\documentclass[final,3p,times]{elsarticle}
\usepackage{amsbsy,amssymb}                     

\journal{Physica C}

\begin{document}

\begin{frontmatter}

\title{Calculation of the phase 
of hidden rotating antiferromagnetic order}
\author{M. Azzouz}
\ead{mazzouz@laurentian.ca}
\address{Department of Physics, Laurentian University,
935 Ramsey Lake Road,
Sudbury, ON P3E 2C6, Canada.}

\begin{abstract}
The phase of the rotating order parameter in rotating 
antiferromagnetism is
calculated using a combination of mean-field theory and
Heisenberg equation. This phase shows a linear
time dependence, which allows us to interpret 
rotating antiferromagnetism as a synchronized 
Larmor-like precession of all the spins in the system or as
an unusual ${\bf q}=(\pi,\pi)$ spin-wave around
a zero local magnetization. 
We discuss 
implications for the 
pseudogap state of high-$T_C$ superconducting materials. 
Rotating antiferromagnetism has been
proposed to model the pseudogap state in these materials.
\end{abstract}
%

\end{frontmatter}

\section{Introduction}

According to several researchers, 
the puzzling pseudogap (PG) phenomenon 
in high-$T_C$ superconductors
(HTSC)  is caused by some sort of hidden order. 
This is
supported by the observation of
a depression in the density of states at 
the Fermi level, with no
order parameter responsible for this depression observed yet
\cite{timusk1998,choiniere2009,chang2010,he2011}. 
Rotating antiferromagnetism (RAF) has been
recently proposed as a possible candidate for 
this hidden order, and several physical 
quantities have already been calculated 
within the RAF theory (RAFT) with good agreement
with available experimental data 
\cite{azzouz2003,azzouz2003b,azzouz2004}. 
RAF is one of several other proposals for the PG 
(see Ref. \cite{he2011} for a discussion). Contrary to 
theories of circulating currents
\cite{varma1997,varma2006,chakravarty2001}, 
RAF is based on the 
concept of an order parameter that has a finite magnitude below 
a critical temperature but a time-dependent phase \cite{azzouz2000}.
Note that all the physical quantities that have so far
been calculated within RAFT
do not depend on the phase of the order parameter in RAF 
\cite{azzouz2003,azzouz2003b,azzouz2004,azzouz2005,azzouz2010,azzouz2010b,azzouz2012}. 
The lack of the time dependence profile for this phase
limited however the full understanding of the nature 
of RAF. The purpose of this work is to calculate this phase 
as a function of time using a combination of RAFT 
and the Heisenberg equation.
We show that it  
varies linearly with time.
As a consequence of this time dependence, 
RAF can be interpreted as a $(\pi,\pi)$ unusual spin wave
around a zero local magnetization or as a synchronized 
Larmor-like precession of all the spins in the system.
Because the phase of this order parameter is time dependent,
it was not possible to calculate it in RAFT alone, 
which is a mean-field approach.

This paper is organized as follows. First in Sec. \ref{sec2.1} 
we rederive RAFT using
the spin ladder operators, which are necessary 
for the phase calculation. In Sec. \ref{sec2.2}, 
we review RAFT.
Then in Sec. \ref{sec2.3} we use the Heisenberg 
equation to get the time dependence for the spin ladder 
operators, which yields the time dependence of the phase 
of the rotating order parameter. In Sec. \ref{sec2.4}, the interpretation of RAF as an unusual $(\pi,\pi)$  
spin wave is explained.  
Finally, conclusions are drawn in Sec. \ref{sec3}.

\section{Approach}

As we are only interested in understanding 
the nature of the PG phase of HTSCs in this work, 
we restrict ourselves to the 
non superconducting phase. 
Consider the $t$-$t'$ Hubbard model in two dimensions:
\begin{eqnarray}
H&=&
-t\sum_{\langle i,j\rangle\sigma}c^{\dag}_{i,\sigma}c_{j,\sigma} 
-t'\sum_{\langle\langle i,j\rangle\rangle\sigma}c^{\dag}_{i,\sigma}c_{j,\sigma}
+ {\rm h.c.} 
\cr
&&-\mu\sum_{i,\sigma}n_{i,\sigma} + U\sum_in_{i\uparrow}
n_{i\downarrow},
\label{hamiltonian1}
\end{eqnarray}
where $\langle i,j\rangle$ and $\langle\langle i,j\rangle\rangle$
designate summation over nearest and second-nearest neighboring sites,
respectively. $t$ and $t'$ are 
electron hopping energies to nearest 
and second-nearest neighbors, 
respectively. Because the phase of RAF is related 
to the spin ladder operators, 
it is useful to rewrite Hamiltonian
(\ref{hamiltonian1}) using these operators.

\subsection{Rewriting the Hamiltonian using the spin ladder operators}
\label{sec2.1}

Using the spin ladder operator written in second quantization
$S_i^+=c^{\dag}_{i,\uparrow}c_{i,\downarrow}$, the onsite Coulomb repulsion
term $Un_{i\uparrow}
n_{i\downarrow}$ can on one hand be cast in the form 
$Un_{i\uparrow}n_{i\downarrow}=Un_{i\uparrow} - US_{i}^+S_i^-$
and on the other hand as
$Un_{i\uparrow}n_{i\downarrow}=Un_{i\downarrow} - US_{i}^-S_i^+$.
Summing and dividing by 2 yields the symmetrized expression
$Un_{i\uparrow}n_{i\downarrow}=\frac{U}{2}(n_{i\uparrow} + n_{i\downarrow}) -\frac{U}{2}(S_{i}^+S_i^- + S_{i}^-S_i^+)$.
The latter can be proved by
calculating the action of each side of the equality on the possible 
states $\{|0\rangle,|\uparrow\rangle,|\downarrow\rangle,
|\uparrow \downarrow\rangle\}$, and noting that 
$S_{i}^+S_i^-|\uparrow \downarrow\rangle
=S_{i}^-S_i^+|\uparrow\downarrow\rangle=0$
due to Pauli exclusion principle, and 
$S_i^+|0\rangle\equiv c_{i\uparrow}^\dag c_{i\downarrow}|0\rangle=0$.
For our many-body system, sites are neither full nor empty, but are 
on average occupied by a density smaller than 1 away from half filling.
Therefore, the terms 
$S_{i}^+S_i^-$ and  $S_{i}^-S_i^+$, which are responsible for
onsite spin-flip excitations, will contribute by lowering 
energy for the sites that are partially occupied by the same density
of spin up and down electrons.
One can decouple 
this term in mean-field theory using $\langle S_i^-\rangle\equiv 
\langle c^{\dag}_{i,\downarrow}c_{i,\uparrow}\rangle$, which leads to a 
collective behavior for the spin-flips, 
and the results obtained in this way are the same as 
in RAFT
\cite{azzouz2003,azzouz2003b,azzouz2004,azzouz2005,azzouz2010,azzouz2010b}.
In this state, a spin flip process at site $i$ 
is simultaneously accompanied by another one at another site $j$; 
the occurrence of the spin flips is synchronized. 
Thermal motion has obviously an effect on this order as it 
does on conventional orders; i.e., above a 
critical temperature (identified with the PG temperature) 
the spin-flip processes become 
uncorrelated, leading to the 
disappearance of the long-range non conventional 
order. 
The spin-flip processes, which 
are purely quantum, continue to exist even above this 
critical temperature, but in an incoherent 
disordered manner.
The occurrence in RAFT of a second-order phase transition at the 
PG temperature is consistent with experimental
data supporting its existence \cite{he2011}.

\subsection{Review of RAFT}
\label{sec2.2}

We rederive
RAFT, which deals with the static part (magnitude) 
of the order parameter $\langle S_i^\pm\rangle$, using the spin ladder operators
then for the dynamic (phase) part we will use the Heisenberg
equation to find its time dependence. To the best of our knowledge
the combination of mean-field 
theory and the Heisenberg equation of quantum mechanics constitutes
a novel approach for the PG in HTSCs.

The parameter 
$Q_i={\langle c_{i,\uparrow}c^{\dag}_{i,\downarrow}\rangle}
=-\langle S_i^-\rangle
\equiv
|Q|e^{i\phi_i}$ is defined in order to carry 
on a mean-field decoupling of the $t$-$t'$ Hubbard model.
Consider the ansatz where $\phi_i-\phi_j=\pi$, 
with $i$ and $j$ labeling any two adjacent
lattice sites. Except for this difference 
of $\pi$ between the phases of the order parameter on two
adjacent sites, the phases $\phi_i\equiv\phi$
are site independent and
assume any value in $[0,2\pi]$.
The normal state Hamiltonian in 
RAFT \cite{azzouz2003,azzouz2003b,azzouz2004} is
\begin{equation}
H\approx\sum_{{\bf k}\in RBZ}\Psi^{\dag}_{\bf k}{\cal H}\Psi_{\bf k}
+NUQ^2-NUn^2,
\label{raft hamiltonian}
\end{equation}
where $N$ is the number of sites, and
$n=\langle n_{i,\sigma}\rangle$ is the expectation value 
of the number operator. 
Because of antiferromagnetic
correlations the lattice consists 
of two sublattices $A$ and $B$, even though 
there is no long-range static antiferromagnetic order.
The summation runs over the reduced
Brillouin zone (RBZ). The Nambu spinor is
$
\Psi^{\dag}_{\bf k}=(c^{A\dag}_{{\bf k}\uparrow}\ c^{B\dag}_{{\bf k}
\uparrow}\ c^{A\dag}_{{\bf k}\downarrow}\ c^{B\dag}_{{\bf k}\downarrow})
$, and the Hamiltonian matrix is 
\[
{\cal H}=
\left( 
\begin{array}{ccccc}
&-\mu'  &\epsilon&Qe^{i\phi}&0 \\ 
&\epsilon & -\mu'&0 & -Qe^{i\phi} \\
&Qe^{-i\phi}&0 &-\mu' &\epsilon \\
&0 & -Qe^{-i\phi}&\epsilon &-\mu'\\
\end{array}
\right),
\]
yielding the energy spectra
$E_{\pm}({\bf k})=-\mu'({\bf k})\pm E_q({\bf k}),
$
where
%
%
$
\mu'({\bf k})=\mu -Un +4t'\cos k_x \cos k_y 
$,
%
%
$E_q({\bf k})=\sqrt{\epsilon^2({\bf k})+(UQ)^2}$, and 
%
%
$
\epsilon({\bf k})=-2t(\cos k_x+\cos k_y)
$.
Because the energy spectra $E_\pm({\bf k})$ 
do not depend on the phase $\phi$ one should be
able to transform $\cal H$ to a matrix that does not depend
on the phase. This can indeed be 
done using the spin-dependent 
gauge transformation
%
%
$
c_{i,\uparrow}\to e^{i\phi/2}c_{i,\uparrow}$
and
$
c_{i,\downarrow}\to e^{-i\phi/2}c_{i,\downarrow}$.
%
%
%
This transformation is equivalent to performing 
a rotation by angle $-\phi$ about the 
$z$ axis for the $x$ and $y$ components of the spin operator.
Indeed, upon using this gauge transformation, 
the spin ladder operators transform according to
%
%
%
$S^+_{i}\to e^{-i\phi}S^+_{i}$ and
$S^-_{i}\to e^{i\phi}S^-_{i}$,
%
%
%
which yields:
\[
\left( 
\begin{array}{cc}
&S_i^x\\ 
&S_i^y
\end{array}
\right)
\to
\left( 
\begin{array}{cc}
\cos\phi  &\sin\phi \\ 
-\sin\phi  &\cos\phi
\end{array}
\right)
\left( 
\begin{array}{cc}
&S_i^x\\ 
&S_i^y
\end{array}
\right).
\]
The thermal averages of $S_i^x$ and $S_i^y$ are given by
\begin{eqnarray}
\frac{\langle S_i^x\rangle}{\hbar} &=& Q\cos\phi,\  
\frac{\langle S_i^y\rangle}{\hbar} = -Q\sin\phi,\ \ 
i \in A, \ {\rm or}\cr
\frac{\langle S_i^x\rangle}{\hbar} &=& -Q\cos\phi,\ 
\frac{\langle S_i^y\rangle}{\hbar} = Q\sin\phi,\ \ 
 i \in B.
\label{mag config}
\end{eqnarray}
Note that $\langle S_i^z\rangle=0$ for $i$ in both sublattices.
Because the phase $\phi$ assumes 
any value
between $0$ and $2\pi$ (see below), rotational symmetry will not look broken
for times greater than the period of rotation.
However if the typical time scale of a probe is much smaller than 
this period symmetry may appear broken. 
\subsection{Calculation of the time dependence of phase $\phi$}
\label{sec2.3}
The magnitude $Q$, which was 
calculated using the minimization of the mean-field 
free energy \cite{azzouz2003,azzouz2003b,azzouz2004},
behaves as in a second-order phase transition in agreement with 
experimental evidence in \cite{he2011}.

Next we calculate the phase using the 
Heisenberg equation 
$\frac{dS_j^+}{d\tau}=\frac{1}{i\hbar}[S_j^+,H]$.
We consider the limit where 
electron hopping is neglected 
in comparison to $\frac{U}{2}(S_j^+S_j^- + S_j^-S_j^+)$. 
The limit considered here is $U\sim 3t$-$5t$; 
this is an intermediate coupling limit where $U>t$ 
but smaller than the bandwidth $\sim8t$ when $t'\ll t$.
It is justified to use this approximation
because spin dynamics is faster than
charge dynamics; i.e., an onsite 
spin flip needs a time $\tau\sim \hbar/U$ to be realized, while 
a charge hopping between adjacent sites 
takes a longer time $\tau\sim\hbar/t$, ($U>t$). In
the Heisenberg equation the undecoupled interaction is used
instead of RAFT's Hamiltonian (\ref{raft hamiltonian})
in order to treat as best as possible quantum fluctuations.
To carry on the calculation, we keep in mind 
that any site $j$ is on average only partially occupied, and that
$|\langle S_j^\pm\rangle|<\hbar/2$. For this reason,
terms like $S_j^+S_j^+S_j^-$ and $S_j^+S_j^-S_j^+$ should be 
kept until the end (these terms normally give zero 
when acting on a spin up state, but a nonzero 
contribution is expected when applied to a partially 
occupied state where 
thermal averages are meaningful and suitable).
In the commutator of the Heisenberg equation
$[S_j^+,H]\approx -\frac{U}{2}[S_j^+,(S_j^+S_j^- + S_j^-S_j^+)]$, we need 
to calculate
$[S_j^+,(S_j^+S_j^- + S_j^-S_j^+)]=[S_j^+,S_j^+S_j^-] 
+[S_j^+,S_j^-S_j^+]=2\hbar(S_j^+S_j^z + S_j^zS_j^+)$.
Using the fundamental
commutation relation $[S_j^z,S_j^+]=\hbar S_j^+$, one gets 
$S_j^+S_j^z + S_j^zS_j^+=\hbar S_j^+ + 2 S_j^+S_j^z$, which leads to
\begin{equation}
\frac{dS_j^+}{d\tau}= iS_j^+\bigg(\frac{U}{\hbar} 
+ \frac{2U}{\hbar^2}S_j^z
\bigg),\ \ \tau\ {\rm is\ time}.
\label{heisenberg eq}
\end{equation}
Again we stress that this equation 
is obtained in the intermediate coupling limit 
($U$ smaller than the bandwidth but higher than 
hopping energies), where
spin dynamics is not governed by the Heisenberg exchange 
coupling $\sim t^2/U$ suitable for the 
strong coupling limit. Eq. (\ref{heisenberg eq}) gives zero when 
acting on state $|\uparrow\rangle$ or $|\downarrow\rangle$. 
However, for a collective state where 
any site is only partially occupied, one has to  
take the thermal average of Eq. (\ref{heisenberg eq}).
One then replaces $S_i^z$ by its RAFT's thermal average,
which is zero. Integrating Eq. (\ref{heisenberg eq}) 
gives for the thermal average
\begin{eqnarray}
\langle S_j^+(\tau)\rangle\approx \langle S_j^+(0)\rangle e^{iU\tau/\hbar},
\label{sol heisenberg eq}
\end{eqnarray}
which yields $\phi=U\tau/\hbar$ modulo $2\pi$
when $\langle S_j^+(0) \rangle$ is identified with  
$|\langle S_j^+(\tau)\rangle|$, 
($-|\langle S_j^+(\tau)\rangle|$), for sublattice $A$, ($B$),
and $e^{i\phi}$ with $e^{iU\tau/\hbar}$. 
The angular frequency is thus
$\omega_{sf}=U/\hbar$, and period 
$T_{sf}=2\pi\hbar/U$ is the time required 
to perform a spin-flip process, or the time needed for 
the rotating order parameter $\langle S_i^{x(y)} \rangle$
to complete a $2\pi$ revolution in a classical point of view. 
The magnetic configuration (\ref{mag config})
takes on the following form 
$
\langle S_i^x\rangle/\hbar = Q\cos(\omega_{sf} \tau)$, 
$\langle S_i^y\rangle/\hbar = -Q\sin(\omega_{sf} \tau)$
for $i$ in sublattice $A$ or
$\langle S_i^x\rangle/\hbar = -Q\cos(\omega_{sf} \tau)$, $\langle S_i^y\rangle/\hbar = Q\sin(\omega_{sf} \tau)$
for $i$ in sublattice $B$, and
$\langle S_i^z\rangle = 0$ 
for $i$ in sublattice $A$ or $B$.

\subsection{Interpretation of rotating antiferromagnetism}
\label{sec2.4}

To illustrate well rotating magnetic order, consider first
the much simpler example of the 
time evolution of a single spin in a magnetic field 
$B$ along the $z$-axis, 
with the initial state given for a spin pointing in the positive 
$x$-direction by
$| S_x,+\rangle = {\frac{1}{\sqrt 2}}(|+\rangle + |-\rangle)$.
The time-dependent
expectation values of the spin components are
$\langle S^x\rangle=\frac{\hbar}{2}\cos(\omega t)$,
$\langle S^y\rangle=\frac{\hbar}{2}\sin(\omega t)$, and 
$\langle S^z\rangle=0$, with $\omega=\frac{|e|B}{m_ec}$. 
$e$ and $m_e$ are the charge and mass of the electron, respectively, 
and $c$ is the speed of light. 
Classically speaking, the spin is confined to rotate about the $z$-axis
in the $xy$ plane with Larmor angular frequency $\omega$.
A rotating ferromagnetic 
state can be realized by 
placing $N$ such states with the same 
frequency on a lattice made of $N$ sites.
For a rotating antiferromagetic state, opposite initial 
states ($\pm| S_x,+\rangle$:
spins point in opposite directions on the $x$-axis) 
are required on each two adjacent sites of the lattice. 
To relate RAF to spin flip processes, we note 
that $\langle S^\pm\rangle=\langle S^x\rangle \pm i\langle S^y\rangle=\frac{\hbar}{2}e^{\pm i\omega t}$ 
in this example. Note that in this example 
is model independent, which may indicate that
all model parameters will do is changing 
multiplying physical factors, 
not the physics itself. In a given model, a coupling 
is necessary for providing the building bloc for RAF, which 
is the precession of a spin (with no local magnetization) 
for each lattice site. The RAF state constructed in this way 
shows a hidden order that can be realized even at finite 
temperature without violating the Mermin-Wagner theorem
\cite{mermin1966}.

The example above allows us to 
interpret RAF as 
a state where spins precess collectively 
in a synchronized manner 
in the spins' $xy$ plane
around an effective
staggered magnetic field $B=m_ecU/\hbar|e|$ 
caused by onsite Coulomb repulsion.
For our many-body system, 
$\hbar/2$ in $\langle S^\pm\rangle=\frac{\hbar}{2}e^{\pm i\omega t}$
is replaced by the magnitude of the
RAF order parameter $Q$, which
can assume
values smaller than $1/2$ due to thermal averaging.
This state is strongly doping dependent. When doping increases,
$Q$ rapidly decreases then vanishes at a doping identified 
as the quantum critical point underneath the superconducting dome
\cite{azzouz2003,azzouz2003b,azzouz2004,azzouz2005,azzouz2010,azzouz2010b}.
In comparison to ordinary spin waves in an antiferromagnet, RAF's state could be viewed
as a ${\bf q}=(\pi,\pi)$ spin wave in an antiferromagnet with zero magnetization. 
Note however that for our system (where $\langle S_i^z\rangle=0$), 
spin-wave theory is not applicable because the spin-wave theory
is built around a stable nonzero
$\langle S_i^z\rangle$ state.

\section{conclusion}
\label{sec3}

The rotating antiferromagnetism theory and 
Heisenberg equation are combined in order to
calculate the phase of the rotating order parameter. 
This phase behaves linearly in time. 
This allows us to interpret rotating
antiferromagnetism in terms of a Larmor-like 
spin precession about an effective magnetic field, 
which is proportional to 
onsite Coulomb repulsion. Another way to see 
rotating antiferromagnetism 
is as an unusual spin-wave at ${\bf q}=(\pi,\pi)$
around a zero magnetization. This work was necessary 
for unveiling the nature of rotating antiferromagnetism, 
which has been proposed for explaining the 
pseudogap behavior in high-$T_C$ materials. Rotating antiferromagnetic order is an example of hidden order, 
which is a serious candidate for the PG state in HTSCs. This
is supported by
the good success of the rotating antiferromagnetism theory
in the calculation of thermodynamics
\cite{azzouz2003,azzouz2003b,azzouz2004}, optical conductivity
\cite{azzouz2005,azzouz2012}, Raman \cite{azzouz2010}, and
angle-resolved photoemission spectroscopy
properties \cite{azzouz2010b}.

Author wishes to thank A.-M. S. Tremblay for 
helpful comments on the manuscript.

\end{document}